\documentclass{iopart}

\usepackage{iopams} 
\usepackage{amstext}

\usepackage{overpic,graphicx}% Include figure files
\usepackage{dcolumn}% eqnarray table columns on decimal point
\usepackage{mathrsfs}
\usepackage{bm}
\usepackage{color,xspace}
\usepackage[small,compact,raggedright]{titlesec}
\usepackage{soul,hyperref}
\usepackage{iopams}
\usepackage{setstack}
\usepackage{graphicx}
\usepackage{xspace}

\newcommand{\eqref}[1]{{(\ref{#1})}}

\newcommand{\uin}[1]{{u_{{#1}}^{\text{in}}}}
\newcommand{\uout}[1]{{u_{{#1}}^{\text{out}}}}
\newcommand{\uh}[1]{{u_{{#1}}^{{\text{hor}}}}}
\newcommand{\ain}[1]{{a_{{#1}}^{\text{in}}}}
\newcommand{\aout}[1]{{a_{{#1}}^{\text{out}}}}
\newcommand{\ket}[1]{\left| {#1} \right\rangle}
\newcommand{\bra}[1]{\left\langle {#1} \right|}
\newcommand{\ah}[1]{{a_{{#1}}^{{\text{hor}}}}}

\newcommand{\proj}[2]{\left| {#1} \right\rangle\!\left\langle {#2} \right|}

\newcommand{\ii}{\mathrm{i}}

\def\b{\begin{equation}}
\def\e{\end{equation}}

\newenvironment{align}{\begin{eqnarray}}{\end{eqnarray}}

\newcommand{\eo}{\eta_{\text{out}}}
\newcommand{\diff}{\text{d}}

\begin{document}

\title{The fate of non-trivial entanglement under gravitational collapse}

\author{Eduardo Mart\'in-Mart\'inez}

\address{Institute for Quantum Computing, Department of Physics and Astronomy, and Department of Applied Mathematics, University of Waterloo, 200 University
Avenue W, Waterloo, Ontario, N2L 3G1, Canada}

\author{Luis J. Garay}

\address{Departamento de F\'isica Te\'orica II, Universidad Complutense de Madrid, 28040 Madrid, Spain}
\address{Instituto de Estructura de la Materia, CSIC, Serrano 121, 28006 Madrid, Spain}
\author{Juan Le\'on}

\address{Instituto de F\'{i}sica Fundamental, CSIC, Serrano 113-B, 28006 Madrid, Spain}

\begin{abstract}
We analyse the evolution of the entanglement of a non-trivial initial quantum field state (which, for simplicity, has been taken to be a bipartite state made out of vacuum and the first excited state)  when it undergoes a gravitational collapse.  We carry out this analysis by generalising the  tools developed to study entanglement behaviour in stationary scenarios and making them suitable to deal with dynamical spacetimes. We  also discuss what kind of problems can be tackled using the formalism spelled out here as well as single out future avenues of research.
\end{abstract}


\maketitle
%%%%%%%%%%%%%%%%%%%%%%%%%%%%%%%%%%
\section{Introduction}
%%%%%%%%%%%%%%%%%%%%%%%%%%%%%%%%%%
The question of how entanglement behaves in non-inertial frames and in curved spacetimes has been around for already a fairly long time. There are many works that centre in the study of uniformly accelerated observers (among many others \cite{Alsingtelep,Alicefalls,AlsingSchul,Edu2,Edu3,Edu4,beyond,Friis2011,CIvyty2,CIvyty3}), or in the background of a stationary eternal black hole \cite{Kerr}. There are also some studies involving entanglement dynamics in
expanding universe  scenarios which have shown that the interaction with the
gravitational field can produce entanglement between quantum field
modes \cite{caball,colapse}.

Focusing on the problem of gravitational collapse, previous works in the literature analysed the correlations between the outgoing and infalling modes in a gravitational collapse when the initial state is the vacuum (see for example \cite{Balbinot,NavarroSalas,BalbinotII,serena,colapse} again among many others). 

In this work we consider  the following  more involved but central issue as far as the behaviour of entanglement in a dynamical spacetime is concerned. In the asymptotic past the field lives in a flat spacetime and its state has some degree of quantum entanglement between two of its modes. Then at some point, gravitational collapse occurs. The collapse makes the observers of the field  unable to access the full state due to the formation of an event horizon. This has an impact on the entanglement that any observer of the field state can acknowledge.

Studying this sort of problems is interesting from many perspectives apart from understanding how quantum correlations behave in dynamical curved spacetimes. Quantum entanglement  plays a key role in black
hole thermodynamics  and the fate of information  in the presence of
horizons. Also, the study of the behaviour of  non-trivial  quantum entanglement in gravitational collapse may be useful for analog gravity proposals that aim at making use of this entanglement as a resource to check genuinely quantum effects derived form the formation of a horizon \cite{serena}.
In general,  this will arguably constitute a rather difficult exercise. However, inspired by tools developed to study the effect of accelerations on quantum entanglement, it may be possible to shed some light on this problem. 

In the study of quantum entanglement from non-inertial perspectives, i.e. in the context of relativistic quantum information, it was not until relatively recently that the physical meaning of the so-called `single mode approximation' was analysed in detail. This approximation was introduced in \cite{Alsingtelep,AlsingMcmhMil}. It consisted in assuming that the Bogoliubov transformations between Minkowski and Rindler modes did not mix frequencies. In 2010 appropriate procedures to construct inertial modes which transform to monochromatic Rindler modes were introduced  in the context of Relativistic Quantum Information  for the accelerated scenario \cite{beyond} as well as in the stationary Schwarzschild scenario \cite{Kerr}.Nowadays we are taking steps towards the analysis of localised states  \cite{drago1} and entanglement behaviour in modes contained in cavities \cite{CIvyty2,CIvyty3}. Although  there are still a number of open questions about the analysis of localised field states and their possible experimental implementability, the two milestones \cite{Alsingtelep,AlsingMcmhMil} and \cite{beyond} have enabled us to understand better the way in which entanglement behaves from non-inertial perspectives. And so, we analyse here the fundamental and qualitative effect of a dynamical gravitational collapse on bipartite entanglement contained in non-trivial quantum field states (that involve vacuum and excited states) prior to the collapse.

In Sec. 2 we introduce the basic spacetime and quantum-field ingredients and tools to analyse the fate of entanglement in a gravitational collapse scenario. Section 3 is devoted to the study of the evolution of the entanglement of a specific bipartite quantum field state made out of vacuum and an excited state (to our knowledge, this is the first time that this kind of non-trivial entanglement in a dynamical spacetime is analysed). Section 4 contains the conclusion and some lines of future research in this context.

\section{Gravitational Collapse}

We will consider a certain maximally entangled state of two modes of the field. This entangled state lives in a spacetime that is originally flat. At some point, a perturbation is produced causing the spacetime to undergo a process of gravitational collapse.
This scenario would very well describe the process of an astrophysical stellar collapse: Prior to the collapse the density of a star is small enough to consider that the spacetime is approximately flat. At some point, the internal forces that kept the star from collapsing fail to counter the gravitational interaction and the star collapses. If nothing stops the collapse, it will reach a point in which an event horizon is formed. 

%If the transition time between the approximately flat to a black hole background is fast enough, 

\begin{figure}
	\begin{center}
		\begin{overpic}[width=.4\columnwidth]{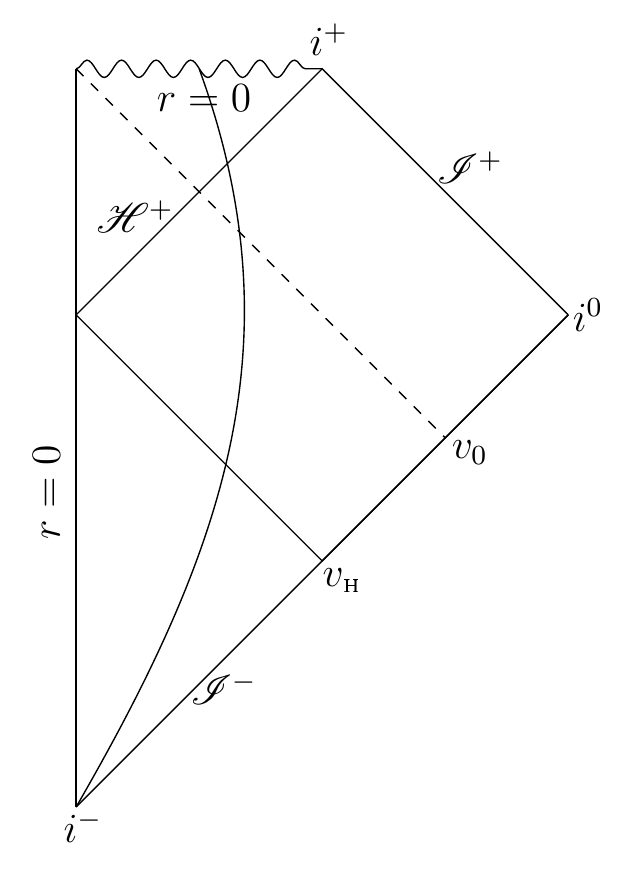}
			\put(20,27){$\longleftarrow$ Stellar collapse}
			\put(45,56){$\longleftarrow$ Radiation shock-wave}
		\end{overpic}
		\caption{Carter-Penrose diagrams for gravitational collapse: Stellar collapse (solid line)
			and ingoing radiation shock-wave (dashed line) giving rise to Vaidya spacetime.}
			\label{fig:vaidya}
		\end{center}
	\end{figure}

Let us consider  the following metric written in terms of ingoing Eddington-Finkelstein  coordinates  as
\begin{equation}
\diff s^2=-\left(1-\frac{2M(v)}{r}\right)\diff v^2+2\diff v\diff r+r^2 \,
\diff \Omega^2,
\end{equation}
where $r$ is the radial coordinate, $v$ is the ingoing null coordinate, and
$M(v)=m\theta(v-v_0)$. For $v_0<v$ this is nothing but the ingoing
Eddington-Finkelstein representation for the Schwarzschild metric
whereas for $v<v_0$ it is just Minkowski spacetime. This metric
represents a radial  ingoing collapsing shockwave of radiation and it is called Vaidya metric  (described schematically in Fig. \ref{fig:vaidya}). This metric is a solution to the
Einstein equations (see for instance. Ref. \cite{NavarroSalas})  that, in spite of its
simplicity, describes very well the gravitational collapse scenario that we want to analyse.  Refinements
of the model to make it more realistic only introduce subleading
corrections. In particular this model captures, up to subleading corrections, the more realistic collapse of a matter cloud. Let $v_{\textsc{h}}=v_0-4m$ be the coordinate of the
last null ray that escapes to the future null infinity $\mathscr{I}^+$
and hence that will eventually form the event horizon (see Fig. \ref{fig:vaidya}).

Consider now a state of a scalar quantum field. We need to introduce convenient bases  of solutions to the
Klein-Gordon equation determined by their behaviour  in the different relevant regions of this collapsing spacetime.  For this,  we will follow a standard procedure (see e.g. \cite{NavarroSalas}).

We first define  the `in'  basis of ingoing positive frequency
modes, associated with the time parameter $v$ at the null past
infinity $\mathscr{I}^-$:
\begin{equation}
\uin{\omega}\sim (4\pi r\sqrt{\omega})^{-1}e^{-\ii\omega v}, \qquad \text{at $\mathscr{I}^-$}.
\end{equation}

Second, we  define another basis in a Cauchy surface in
the future. In this case, the asymptotic future
$\mathscr{I}^+$ is not a Cauchy surface in itself, so we need to  consider also the future event horizon
$\mathscr{H}^+$. Let us begin with the `out' modes defined as being outgoing
positive-frequency in terms  of the natural time parameter
$\eta_{\text{out}}$ at  $\mathscr{I}^+$, which are
\begin{equation}
\uout{\omega}\sim(4\pi r\sqrt{\omega})^{-1}e^{-\ii\omega \eta_{\text{out}}},\qquad \text{at $\mathscr{I}^+$},
\end{equation}
where  $\eo=v-2r^*_{\text{out}}$ and $r^*_{\text{out}}
%=r_{\text{out}}+2m\log|1-r_{\text{out}}/(2m)|
$ is the radial tortoise coordinate in the Schwarzschild region. At early  times, these modes $\uout{\omega}$ concentrate near $v_{\textsc{h}}$ at $\mathscr{I}^-$ and behave in the following way:
\begin{equation}\label{uout2}
\uout{\omega}\approx(4\pi r\sqrt{\omega})^{-1}
e^{-\ii\omega\left(v_{\textsc{h}}-4m\ln\frac{|v_{\textsc{h}}-v|}{4m}\right)}
\theta(v_{\textsc{h}}-v),\qquad \text{at $\mathscr{I}^-$},
\end{equation}
having support only in the region $v < v_{\textsc{h}}$, since only the rays of light that depart from $v < v_{\textsc{h}}$ will
reach the asymptotic region $\mathscr{I}^+$. The rest will fall down into
the horizon. 

Finally, we use an analytical continuation argument to define the `hor' modes at  $\mathscr{H}^+$: These will be modes that behave as $\uout{\omega}$ in the asymptotic past $\mathscr{I}^-$, but for $v > v_{\textsc{h}}$. In other words, we define them as modes that leave the asymptotic past to fall into the horizon, never reaching the asymptotic future. Near the Cauchy surface $\mathscr{I}^-$, these modes behave as
\begin{equation}
\uh{\omega}\sim(4\pi r\sqrt{\omega})^{-1}
e^{+\ii\omega\left(v_{\textsc{h}}-4m\ln\frac{|v_{\textsc{h}}-v|}{4m}\right)}
\theta(v-v_{\textsc{h}}),\qquad \text{at $\mathscr{I}^-$}.
\end{equation}

By expanding the field in terms of the two sets of modes (`in' on the one hand and  `hor-out' on the other) we can relate the two sets of solutions via the corresponding Bogoliubov coefficients. We will provide more details below, but for now let us refer to  the extensive literature on this topic (see for instance \cite{NavarroSalas}) and directly give the expression of the bosonic annihilation field operators in the asymptotic past  in terms of the creation and annihilation operators of  `out' and  `hor' modes which, in the notation of  \cite{colapse}, read:
\begin{equation}
\nonumber\ain{\omega'}=\int d\omega \Big[\alpha^*_{\omega\omega'}
\big(\aout{\omega}-\tanh r_{\omega}\,\ah{\omega}^\dagger\big)+\alpha_{\omega\omega'}e^{i\varphi}\big(\ah{\omega}-\tanh r_{\omega}\,
\aout{\omega}^\dagger\big)\Big],
\label{ain}
\end{equation}
where  $\tanh r_{\omega} = e^{-4\pi m\omega}$. The 
values of $\varphi$ and  $\alpha_{\omega\omega'}$ will be given below.
The `in' vacuum, given by $\ain{\omega}\ket{0}_{\text{in}}=0$  for all positive frequencies $\omega$, can be readily rewritten  in the `out-hor' basis as
\begin{equation}\label{sque}
\ket{0}_{\text{in}}=\prod_{\omega}\frac{1}{\cosh r_{\omega}}
\sum_{n=0}^\infty (\tanh r_{\omega})^{n}\ket{n_{\omega}}_{\text{hor}}
\ket{n_{\omega}}_{\text{out}},
\end{equation}
where $\ket{n_{\omega}}$ denotes a mode with occupation number $n$ and  frequency $\omega$.

The next task is to write the one-particle state in the past $\ket{1_\omega}_{\text{in}}$  as a linear combination of the `out-hor' basis modes. If we have a monochromatic excitation in the asymptotic past, equation \eqref{ain} tells us that it will become a highly non-monochromatic linear combination of `hor' and `out' modes. 
The standard well-known procedure (which we briefly summarise in what follows, see e.g. \cite{NavarroSalas,beyond}) is to construct another basis of  `in' modes with positive-frequency in the past such that its Bogoliubov transformation into `hor' and `out' modes is diagonal in frequencies. Let us call those modes $u_{\Omega}^{\text{\text{R}}}$ and $u_{\Omega}^{\text{\text{L}}}$ in order to keep the notation of Ref.~\cite{beyond}. Note  that, for all the reasons discussed above, these modes are intrinsically non-monochromatic in the asymptotic past, i.e. in the basis of modes $\uin{\omega}$, and that $\Omega$ is labelling the frequency of such modes $u_{\Omega}^{\text{\text{R}}}$ and $u_{\Omega}^{\text{\text{L}}}$ with respect to the time in the asymptotic future region `out'.

As suggested by the fact that   $\ket0_{\text{in}}$ is a two-mode squeezed vacuum   of `out' and `hor' modes, let us define new positive-norm `R-L'  modes  by the following  diagonal Bogoliubov  transformation from `out-hor' basis with the form of a two-mode squeezing operation:
\begin{eqnarray}
\label{eq:unruhmodes}
u_{\Omega}^{\text{\text{R}}} &=& 
\cosh(r_\Omega)  u_{\Omega}^{\text{out}} + \sinh(r_\Omega)  {u_{\Omega}^{\text{hor}}}^{\ast}, \nonumber\\
u_{\Omega}^{\text{\text{L}}} &=& 
\cosh(r_\Omega)  u_{\Omega}^{\text{hor}} + \sinh(r_\Omega) {u_{\Omega}^{\text{out}}}^{\ast}.
\end{eqnarray} 
Taking Klein-Gordon inner products  with the `in' modes we find the following form  of $u_{\Omega}^{\text{\text{R}}}$ and $u_{\Omega}^{\text{\text{L}}}$ in terms of the   $\uin\omega$ modes:
\begin{align}
u_{\Omega}^{\text{\text{R}}}&= \int_{0}^{\infty} [\cosh(r_\Omega)\alpha_{\omega\Omega}+\sinh(r_\Omega)\delta_{\omega\Omega}]^\ast u_{\omega}^{\text{in}} \text{d}\omega,\\
u_{\Omega}^{\text{\text{L}}}&= \int_{0}^{\infty} [\cosh(r_\Omega)\gamma_{\omega\Omega}+\sinh(r_\Omega)\beta_{\omega\Omega}]^\ast u_{\omega}^{\text{in}} \text{d}\omega,
\label{Unruhmodes}
\end{align} 
$\!\!$where $\alpha_{\omega\Omega}=\left(u_{\omega}^{\text{in}}, u_{\Omega}^{\text{out}}\right),\ \beta_{\omega\Omega}=\big(u_{\omega}^{\text{in}},{u_{\Omega}^{\text{out}}}^{\ast}\big),\ \gamma_{\omega\Omega}=\left(u_{\omega}^{\text{in}}, u_{\Omega}^{\text{hor}}\right)$ and $\delta_{\omega\Omega} =\big(u_{\omega}^{\text{in}},{u_{\Omega}^{\text{hor}}}^{\ast}\big)$ are the following Bogoliubov coefficients
\begin{equation}
\alpha_{\omega\Omega}=\frac{-1}{2\pi}\sqrt{\frac{\Omega}{\omega}}(4m)^{-4\ii m\omega}e^{-\ii(\omega-\Omega)v_{\textsc{h}}}(-\ii\Omega)^{-1-4\ii m\omega}\,\Gamma(1+4\ii m\omega)
\label{bogos1}
\end{equation}
\begin{equation}
\beta_{\omega\Omega}=-\tanh r_\omega\, e^{-2\ii\Omega v_{\textsc{h}} }\alpha_{\omega\Omega}\qquad \gamma_{\omega\Omega}=e^{-2\ii\Omega v_{\textsc{h}} }\alpha_{\omega\Omega}\qquad \delta_{\omega\Omega} =-\tanh r_\omega \alpha_{\omega\Omega}^{\ast}.
\label{bogos2}
\end{equation}
Therefore we see  that the `R-L' modes  are purely positive frequency linear combinations of the `in' modes  and that they also form a complete set of solutions of the field equation in the asymptotic past.

This relationship between the modes directly translates into a relation between the particle operators associated with them
\begin{align}
{a_{\Omega}^{\text{\text{R}}}}=\cosh r_\Omega \, a_{\Omega}^{\text{out}}-\sinh r_\Omega \,  {a_{\Omega}^{\text{hor}}}^\dagger,\\
{a_{\Omega}^{\text{\text{L}}}}=\cosh r_\Omega \, a_{\Omega}^{\text{hor}}-\sinh r_\Omega \,  {a_{\Omega}^{\text{out}}}^\dagger.
\end{align}
Obviously, these operators annihilate the `in' vacuum:\quad ${a_{\Omega}^{\text{\text{R}}}}\ket{0}_{\text{in}}=
{a_{\Omega}^{\text{\text{L}}}}\ket{0}_{\text{in}}=0$.

To summarise, these modes have the following features:
\begin{enumerate}
	\item They share the same vacuum $\ket{0}_{\text{in}}$ as the monochromatic modes $\ain{\omega}$.
	\item They form a complete basis of solutions to the field equation which are positive-frequency in the asymptotic past.
	\item They translate into a single frequency mode when expressed in the future basis.
\end{enumerate}
Properties (i) and (ii) allow us to decompose any physical state as some combination of these modes, which makes them worth studying as an intermediate stage of more general cases. The third feature (diagonal Bogoliubov transformations) greatly simplifies the formalism, enabling us to use all the artillery already deployed in other simpler scenarios also in the case of stellar collapse, providing us with a nice and clear interpretation to the analysis of the entanglement in the asymptotic future.

Repeating an analogous reasoning as in \cite{beyond}, we can still introduce a more general annihilation operator with this properties, which will be a linear combination of the two annihilation operators $a_{\Omega}^{\text{\text{R}}}$ and $a_{\Omega}^{\text{\text{L}}}$ defined above:
\begin{equation}
C_{\Omega,q_{\text{R}}}= q_{\text{R}}{a_{\Omega}^{\text{\text{R}}}}+q_{\text{L}} {a_{\Omega}^{\text{\text{L}}}}\label{runn},\label{genmode}\end{equation}
where $q_{\text{L}}$ and $q_{\text{R}}$ are real parameters satisfying $ q_{\text{L}}=\sqrt{1-  q_{\text{R}} ^2}$ and $2^{-1/2}\leq q_{\text{R}} \leq 1$. 

%As we will discuss below, introducing these modes will allow us to study how entanglement behaves in the presence of a stellar collapse.

\section{Entanglement behaviour}

Let us consider the following maximally entangled bipartite state in the asymptotic past, ``prepared'' long before  collapse starts
\begin{equation}\label{maxent}
\ket{\Psi}=\frac{1}{\sqrt2}\left(\ket{0}^{\text{A}}_{\text{in}}\ket{0}^{\text{B}}_{\text{in}}+\ket{1}^{\text{A}}_{\text{in}}\ket{1_\Omega}^{\text{B}}_{\text{in}}\right),
\end{equation}
where the excited modes for Bob are chosen  to be those generated by \eqref{runn}, namely,
 \begin{equation}
\ket{1_\Omega}_{\text{in}}=C^\dagger_{\Omega,q_{\text{R}}}\ket{0}_{\text{in}},
\end{equation}
 whereas Alice's mode can be chosen arbitrarily (that is why it is not labelled with $\Omega$). This initial state \eqref{maxent} will be observed by two observers Alice and Bob. While we will consider that Alice  has unrestricted access to her partial state, we will assume that Bob does not  because at some point the process of gravitational collapse will generate an event horizon, preventing him from accessing the full state.

Although perhaps the most natural scenario would be that in which both subsystems are in the proximities of a stellar collapse and, hence, both observers would undergo similar processes, let us consider for simplicity that only one of the subsystems is going to be affected by the stellar collapse. One can think of Alice's state prepared such that  it is a localized state living far away from the collapsing star.  Alternatively we could consider that Alice can measure her subsystem prior to the formation of the horizon so that it cannot hinder her ability to obtain information about her partial state. 
On the other hand, Bob's knowledge about his subsystem is going to be limited because between the time when the state was created and the time in which he will be able to measure it, an event horizon  appears, preventing him from accessing the full state in the future. This scenario is not devoid of physical interest. As we will show later on, it will allow us to focus on questions regarding quantum correlations between modes in the past and modes falling into the horizon.

In these circumstances, Alice measures in the `in' basis whereas Bob  measures in the `out' basis, having lost all the information contained in the modes `hor' that are bound to fall into the forming black hole. 
Now, to describe the effective state to which Alice and Bob can access requires that  we trace out the modes that become causally disconnected from Bob due to the formation of the horizon, i.e. the `hor' modes:
\begin{equation}
\rho_{A-\text{out}}=\tr_{\text{hor}}(\proj{\Psi}{\Psi}).
\end{equation}
This state is non-separable and we can compute its negativity \cite{Negat} as a convenient quantifier of quantum entanglement.
In simple words, when we compute the negativity of the density matrix $\rho_{A-\text{out}}$ we obtain a measurement of the correlations between Alice's fully accessible state and the modes that reach the asymptotic future escaping the collapse.

It is legit then to ask about what would be the quantum entanglement between Alice's state and those modes that will not make it to the asymptotic future because they will fall into the incipient horizon, becoming trapped into the black hole. The state whose separability we would have to analyse  would then be
\begin{equation}
\rho_{A-\text{hor}}=\tr_{\text{out}}(\proj{\Psi}{\Psi}).
\end{equation}

The procedure to compute the negativity is the following: first we express $\ket{\Psi}$ in the basis of `hor-out' modes for Bob, which yields
\begin{align}
\ket{\Psi}&=\frac{1}{\sqrt2}\left[\frac{1}{\cosh r_\Omega} \sum_{n=0} \tanh^n r_\Omega \ket{0}^{\text{A}}_{\text{in}}\ket{n_\Omega}^{\text{B}}_{\text{out}}
\ket{n_\Omega}^{\text{B}}_{\text{hor}}\right.\nonumber\\
&\left. +\frac{1}{\cosh^2 r_\Omega} \sum_{n=0}\sqrt{n+1} \tanh^n r_\Omega \ket{1}^{\text{A}}_{\text{in}}(q_{\text{R}}\ket{(n+1)_\Omega}^{\text{B}}_{\text{out}}
\ket{n_\Omega}^{\text{B}}_{\text{hor}}+
q_{\text{L}}\ket{n_\Omega}^{\text{B}}_{\text{out}}
\ket{(n+1)_\Omega}^{\text{B}}_{\text{hor}}) \right].
\label{equs}
\end{align}

Then the density matrices $\rho_{A-\text{hor}}$ and $\rho_{A-\text{out}}$ are obtained after a simple but lengthy algebra exercise by tracing out the `out' and `hor' modes respectively from $\ket{\Psi}$,

\begin{eqnarray}
\rho_{A-\text{out}}=\sum_{n=0} \frac{T_r^{2n}}{2C_r^2}
\ket{0}^{\text{A}}_{\text{in}}\ket{n_\Omega}^{\text{B}}_{\text{out}}
\bra{0}^{\text{A}}_{\text{in}}\bra{n_\Omega}^{\text{B}}_{\text{out}} \nonumber\\
+\sum_{n=0} (n+1) \frac{T_r^{2n}}{2C_r^4}
\left(q_{\text{R}}^2 \ket{1}^{\text{A}}_{\text{in}}\ket{(n+1)_\Omega}^{\text{B}}_{\text{out}}
\bra{1}^{\text{A}}_{\text{in}}\bra{(n+1)_\Omega}^{\text{B}}_{\text{out}}+
q_{\text{L}}^2 \ket{1}^{\text{A}}_{\text{in}}\ket{n_\Omega}^{\text{B}}_{\text{out}}
\bra{1}^{\text{A}}_{\text{in}}\bra{n_\Omega}^{\text{B}}_{\text{out}}\right) \nonumber\\
+\sum_{n=0}\sqrt{(n+1)(n+2)} \frac{T_r^{2n+1}}{2C_r^4}\,q_{\text{R}}q_{\text{L}}
\left(\ket{1}^{\text{A}}_{\text{in}}\ket{n_\Omega}^{\text{B}}_{\text{out}}
\bra{1}^{\text{A}}_{\text{in}}\bra{(n+2)_\Omega}^{\text{B}}_{\text{out}}+
\ket{1}^{\text{A}}_{\text{in}}\ket{(n+2)_\Omega}^{\text{B}}_{\text{out}}
\bra{1}^{\text{A}}_{\text{in}}\bra{n_\Omega}^{\text{B}}_{\text{out}}\right)\nonumber\\
+\sum_{n=0}\sqrt{n+1} \frac{T_r^{2n+1}}{2C_r^3}\,\left[q_{\text{R}}
\left(\ket{0}^{\text{A}}_{\text{in}}\ket{n_\Omega}^{\text{B}}_{\text{out}}
\bra{1}^{\text{A}}_{\text{in}}\bra{(n+1)_\Omega}^{\text{B}}_{\text{out}}+
\ket{1}^{\text{A}}_{\text{in}}\ket{(n+1)_\Omega}^{\text{B}}_{\text{out}}
\bra{0}^{\text{A}}_{\text{in}}\bra{n_\Omega}^{\text{B}}_{\text{out}}\right)\right.\nonumber\\
+\left.q_{\text{L}}\left(\ket{0}^{\text{A}}_{\text{in}}\ket{(n+1)_\Omega}^{\text{B}}_{\text{out}}
\bra{1}^{\text{A}}_{\text{in}}\bra{n_\Omega}^{\text{B}}_{\text{out}}+
\ket{1}^{\text{A}}_{\text{in}}\ket{n_\Omega}^{\text{B}}_{\text{out}}
\bra{0}^{\text{A}}_{\text{in}}\bra{(n+1)_\Omega}^{\text{B}}_{\text{out}}\right)\right],
\end{eqnarray}
where $T_r=\tanh r_\Omega$ and $C_r=\cosh r_\Omega$. The expression for $\rho_{A-\text{hor}}$ is obtained  by exchanging $q_R$ and $q_L$ and `out' by `hor' in the equation above.

The entanglement monotone that we will compute, the negativity \cite{Negat}, is the sum of the negative eigenvalues of the partial transposed density matrix of the quantum state for which we want to evaluate its degree of distillable entanglement. To compute it we first take partial transposes in $\rho_{A-\text{out}}$ and $\rho_{A-\text{hor}}$ (which is the transpose only with respect to Alice's indices). This yields
\begin{eqnarray}
\rho^{T_{\text{A}}}_{A-\text{out}}=\sum_{n=0} \frac{T_r^{2n}}{2C_r^2}
\ket{0}^{\text{A}}_{\text{in}}\ket{n_\Omega}^{\text{B}}_{\text{out}}
\bra{0}^{\text{A}}_{\text{in}}\bra{n_\Omega}^{\text{B}}_{\text{out}} \nonumber\\
+\sum_{n=0} (n+1) \frac{T_r^{2n}}{2C_r^4}
\left(q_{\text{R}}^2 \ket{1}^{\text{A}}_{\text{in}}\ket{(n+1)_\Omega}^{\text{B}}_{\text{out}}
\bra{1}^{\text{A}}_{\text{in}}\bra{(n+1)_\Omega}^{\text{B}}_{\text{out}}+
q_{\text{L}}^2 \ket{1}^{\text{A}}_{\text{in}}\ket{n_\Omega}^{\text{B}}_{\text{out}}
\bra{1}^{\text{A}}_{\text{in}}\bra{n_\Omega}^{\text{B}}_{\text{out}}\right) \nonumber\\
+\sum_{n=0}\sqrt{(n+1)(n+2)} \frac{T_r^{2n+1}}{2C_r^4}\,q_{\text{R}}q_{\text{L}}
\left(\ket{1}^{\text{A}}_{\text{in}}\ket{n_\Omega}^{\text{B}}_{\text{out}}
\bra{1}^{\text{A}}_{\text{in}}\bra{(n+2)_\Omega}^{\text{B}}_{\text{out}}+
\ket{1}^{\text{A}}_{\text{in}}\ket{(n+2)_\Omega}^{\text{B}}_{\text{out}}
\bra{1}^{\text{A}}_{\text{in}}\bra{n_\Omega}^{\text{B}}_{\text{out}}\right)\nonumber\\
+\sum_{n=0}\sqrt{n+1} \frac{T_r^{2n+1}}{2C_r^3}\,\left[q_{\text{R}}
\left(\ket{1}^{\text{A}}_{\text{in}}\ket{n_\Omega}^{\text{B}}_{\text{out}}
\bra{0}^{\text{A}}_{\text{in}}\bra{(n+1)_\Omega}^{\text{B}}_{\text{out}}+
\ket{0}^{\text{A}}_{\text{in}}\ket{(n+1)_\Omega}^{\text{B}}_{\text{out}}
\bra{1}^{\text{A}}_{\text{in}}\bra{n_\Omega}^{\text{B}}_{\text{out}}\right)\right.\nonumber\\
+\left.q_{\text{L}}\left(\ket{1}^{\text{A}}_{\text{in}}\ket{(n+1)_\Omega}^{\text{B}}_{\text{out}}
\bra{0}^{\text{A}}_{\text{in}}\bra{n_\Omega}^{\text{B}}_{\text{out}}+
\ket{0}^{\text{A}}_{\text{in}}\ket{n_\Omega}^{\text{B}}_{\text{out}}
\bra{1}^{\text{A}}_{\text{in}}\bra{(n+1)_\Omega}^{\text{B}}_{\text{out}}\right)\right],
\end{eqnarray}
where, as above, the expression for $\rho^{T_{\text{A}}}_{A-\text{hor}}$ is obtained  by exchanging $q_R$ and $q_L$ and `out' by `hor'. 
 
As it is the case of the accelerated-observer scenario,  the diagonalisation of the infinite-dimensional partial transposed density matrices  $\rho^{T_{\text{A}}}_{A-\text{hor}}$ and $\rho^{T_{\text{A}}}_{A-\text{hor}}$ can be carried out only numerically since, with the exception of the case $q_{\text{R}}=1$, no block-diagonalisation can be  performed.

\begin{figure}[h]
\begin{center}
\includegraphics[width=.9\columnwidth]{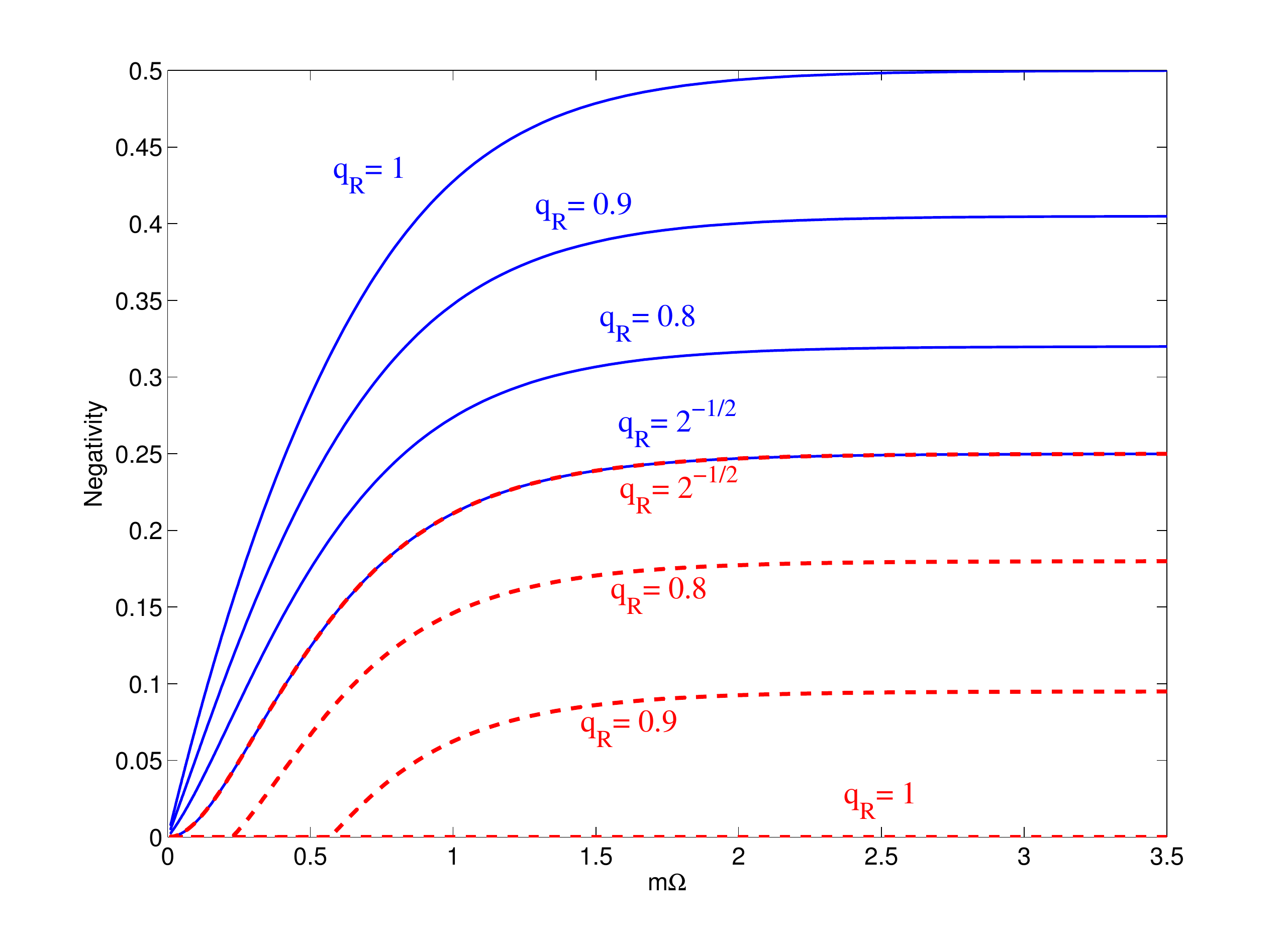}
\caption{Negativity of $\rho_{A-\text{out}}$ (solid blue line) and $\rho_{A-\text{hor}}$ (red dashed line) as a function of the product of mass of the black hole and probed frequency for various choices of the past modes \eqref{genmode}. Note that no entanglement survives the singular black hole limit ($m\rightarrow0$) where the Hawking temperature is divergent.}
\label{fig:plot}
\end{center}
\end{figure}

Figure \ref{fig:plot} shows the result of the calculations. The negativity `A-out' as a function of the mass of the forming black hole and the frequency of the probed `out' mode is shown   as  solid blue lines. The negativity `A-hor' is plotted in the same figure as dashed red lines.

We see that varying the parameter $q_{\text{R}}$ we are basically controlling whether Alice's state will have more quantum correlations with the modes that will reach the asymptotic future or the modes that will fall into the horizon. This is best seen in the large black hole mass limit, where there is an exact trade-off between the correlations that Alice's mode has with infalling and outgoing modes.

We can also see that for very small black holes, all entanglement is completely degraded: when $m\rightarrow 0$ no entanglement survives either with the modes that fall into the horizon or the modes that will reach the future.
As the mass of the black hole (or the frequency of the probed mode) is increased, the correlations quickly become insensitive to the presence of the horizon, as one can expect taking a look at the quantum effects induced by gravity in the presence of a black hole: They become stronger as the mass of the black hole is closer to zero. This suggests that this can be understood in a pictorial  way as a limit in which the Hawking-like radiation spoils all correlations contained in the state. 

As it is well known \cite{Hawking}, if Bob can only measure modes in the asymptotic future, he will see the vacuum state $\ket0_{\text{in}}$ as a thermal state. Indeed, if we compute  how the `in' vacuum is seen by observers in the asymptotic future we obtain that
$\rho^{\ket{0}_{\text{in}}}_{\text{out}}=\tr_{\text{hor}}(\ket{0}_{\text{in}}\!\!\bra{0})=\prod_\omega\rho_{\text{out},\omega}$, where
\begin{equation}
\rho_{\text{out},\omega}=\frac{1}{(\cosh r_{\omega})^2}
\sum_{n=0}^\infty (\tanh r_{\omega})^{2n}
\ket{n_{\omega}}_{\text{out}}\!\!\bra{n_{\omega}}.
\end{equation}
This is a thermal radiation state whose temperature is 
$T_{\text{H}}=(8\pi m)^{-1}$.
So if that happens with the vacuum state it would be a reasonable hand-waving argument that, if instead of the vacuum, we consider a pre-existing non-trivial entangled state such as that of equation \eqref{maxent}, the thermal-like noise could impair the ability of the observers of acknowledging quantum correlations in the system.

However one has to be very careful when thinking to what extent this behaviour can be naively associated to the Hawking thermal noise. To begin with, we are not considering the vacuum state, but rather an entangled  state of field excitations. The process of change of basis and tracing out of the modes that fall into the event horizon is not as trivial as for the vacuum case. In fact,  it has been shown that in the Rindler scenario \cite{MigC}  and beyond the so-called single mode approximation with some choices of the state, the accessible entanglement for an accelerated observer may behave in a non-monotonic way, as opposed to the first results reported in~\cite{Alicefalls,AlsingSchul,beyond}. This is due to inaccessible correlations in the initial states becoming accessible to the accelerated observer when his proper Fock basis changes as acceleration varies. While this phenomenon was highlighted in \cite{MigC} for the Rindler case, for an analogous choice of  the modes \eqref{runn}, a similar behaviour would be expected in the dynamic scenario analysed  here.  

Let us conclude with a note of warning: The modes analysed here as tools to study entanglement behaviour in gravitational collapse  have very nice properties but due to their highly non-monochromatic nature and non-localisation they are modes that can arguably be difficult to prepare and measure in an hypothetical experiment. 
This said, this tool will allow us to simplify the calculations so that we can extract fundamental results in settings in which other techniques have proven not operational, much in the same fashion as the introduction of Unruh modes \cite{beyond} has allowed progress in our understanding of quantum correlations from non-inertial perspectives. One has to keep in mind that the modes used here share some fundamental properties with the standard monochromatic `in' modes, and that they form a complete basis of solutions to the field equations. This means that any physically conceivable state can be expressed as a superposition of modes as the ones studied here. 

\section{Conclusion and Future research}

We have analysed the behaviour of quantum entanglement present in some initial field state when a gravitational collapse occurs in the background. 
The quantum correlations in that state are perturbed  by the formation of the event horizon, mixing the quantum state originally prepared, and therefore, degrading the original correlations.
We have done so by adapting  the tools developed in the analysis of entanglement in the context of the Unruh-Hawking effect \cite{Alicefalls,Kerr} to go beyond what was known as `single mode approximation' \cite{beyond}. 

We have shown, that similarly to the infinite acceleration limit in accelerated scenarios \cite{Alicefalls,Edu4,beyond} and similar to the stationary eternal blackhole scenario \cite{Kerr}, entanglement is completely degraded when we consider singular black holes ($m\rightarrow0$) for which the Hawking temperature diverges. 

A trivial extension of the results obtained  here is considering that the appearance of an event horizon  affects the ability to access the full state for  both Alice and Bob. In these cases, and for maximally entangled states of a scalar field, entanglement will be arguably degraded more quickly than in the case where only Bob is affected by the collapse, much in a similar way as it happens in the acceleration scenario \cite{Adeschul}. Extending this result to the case where both observers measure after the horizon is created is somewhat straightforward with the tools developed here, being mainly a matter of a more complicated calculation, whereas the results are arguably going to be qualitatively the same.

The next natural step is to introduce localised measurements that will endow  the entanglement degradation phenomena reported here with  operational meaning. Using for example, localised projective measurements as in \cite{drago1} more physical scenarios can be analysed.

\section*{Acknowledgements}

We would like to thank Robert B. Mann and Tim Ralph for their kind invitation to contribute to the Focused issue of Class. and Quantum Grav. on Relativistic Quantum Information. This
work was  supported by the Spanish MICINN/MINECO Projects
FIS2011-29287, FIS2008-06078-C03-03, FIS2011-30145-C03-02, the CAM research
consortium QUITEMAD S2009/ESP-1594, and the Consolider-Ingenio
2010 Program CPAN (CSD2007-00042).

\section*{References}

\bibliographystyle{bibstyleTMP}
%\bibliography{references}

\end{document}